\documentclass[11pt,twoside]{article}  
\usepackage{asp2006}
\usepackage{adassconf}

\begin{document}   

%
%

\paperID{P24}

%

\title{Pointing the SOFIA Telescope}

%
%
%
%
%

\markboth{Gross, Rasmussen, and Moore}{Pointing the SOFIA Telescope}

%
%
%
%

\author{Michael A. K. Gross\altaffilmark{1},
        John J. Rasmussen\altaffilmark{2},
        Elizabeth M. Moore\altaffilmark{1}}
\altaffiltext{1}{Universities Space Research Assoc.,
                 NASA/Ames Research Center,
                 Moffett Field, CA, USA}
\altaffiltext{2}{Critical Realm Corp.,
                 San Jose, CA, USA}

%

\contact{Michael A. K. Gross}
\email{mgross@sofia.usra.edu}

%
%
%

\paindex{Gross, M.~A.~K.}
\aindex{Rasmussen, J.~J.}     
\aindex{Moore, E.~M.}     

%



\begin{abstract}          
SOFIA is an airborne, gyroscopically stabilized 2.5m infrared telescope,
mounted to a spherical bearing. Unlike its predecessors, SOFIA will work in
absolute coordinates, despite its continually changing position and attitude.
In order to manage this, SOFIA must relate equatorial and telescope coordinates
using a combination of avionics data and star identification,
manage field rotation and track sky images.
We describe the algorithms and systems
required to acquire and maintain the equatorial reference frame, relate it
to tracking imagers and the science instrument, set up the oscillating
secondary mirror, and aggregate pointings into relocatable
nods and dithers.
\end{abstract}

%
%

\section{SOFIA Telescope-Pointing Architecture}
\begin{figure}[!ht]
\epsscale{0.50}
\plotone{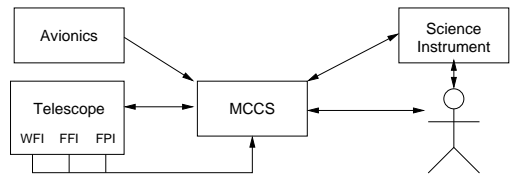}
\caption{The telescope-pointing systems on the SOFIA aircraft.}
\label{fig:sys}
\end{figure}
The SOFIA pointing architecture is shown in Fig.~\ref{fig:sys}.
The Telescope Assembly (TA) keeps track of its attitude using gyroscopes
corrected by tracking imagers, but is unaware of sky
coordinates.  The Science Instrument (SI) is responsible for science data
acquisition.  Avionics provide aircraft location and attitude necessary to
initialize sky coordinates.  The Mission Communications \& Control System
(MCCS) correlates gyroscopes
with sky coordinates and translates pointing building blocks between
coordinate systems.

\section{Coordinate Systems}
The TA describes its attitude as a 3 degree-of-freedom (DOF) rotation from the
telescope
reference frame (TARF) to various coordinate
systems mechanically attached to the telescope.
The MCCS translates these to external coordinates, and accounts for the
optical path.
The TA is far from rigid and orthogonal; thus many intermediate
coordinate systems are established.
Directly-observable MCCS coordinate systems are Equatorial (ERF) and
grid coordinates for the three tracking imagers and the Science Instrument
(SIRF).
Key internal coordinate systems are the telescope and inertial references
(TARF and IRF) with IRF derived from a set of gyroscopes.
The MCCS distributes positions in all coordinates
for real-time monitoring, FITS headers, and archiving.

\section{Establishing Equatorial Coordinates}
\begin{figure}[!ht]
\plottwo{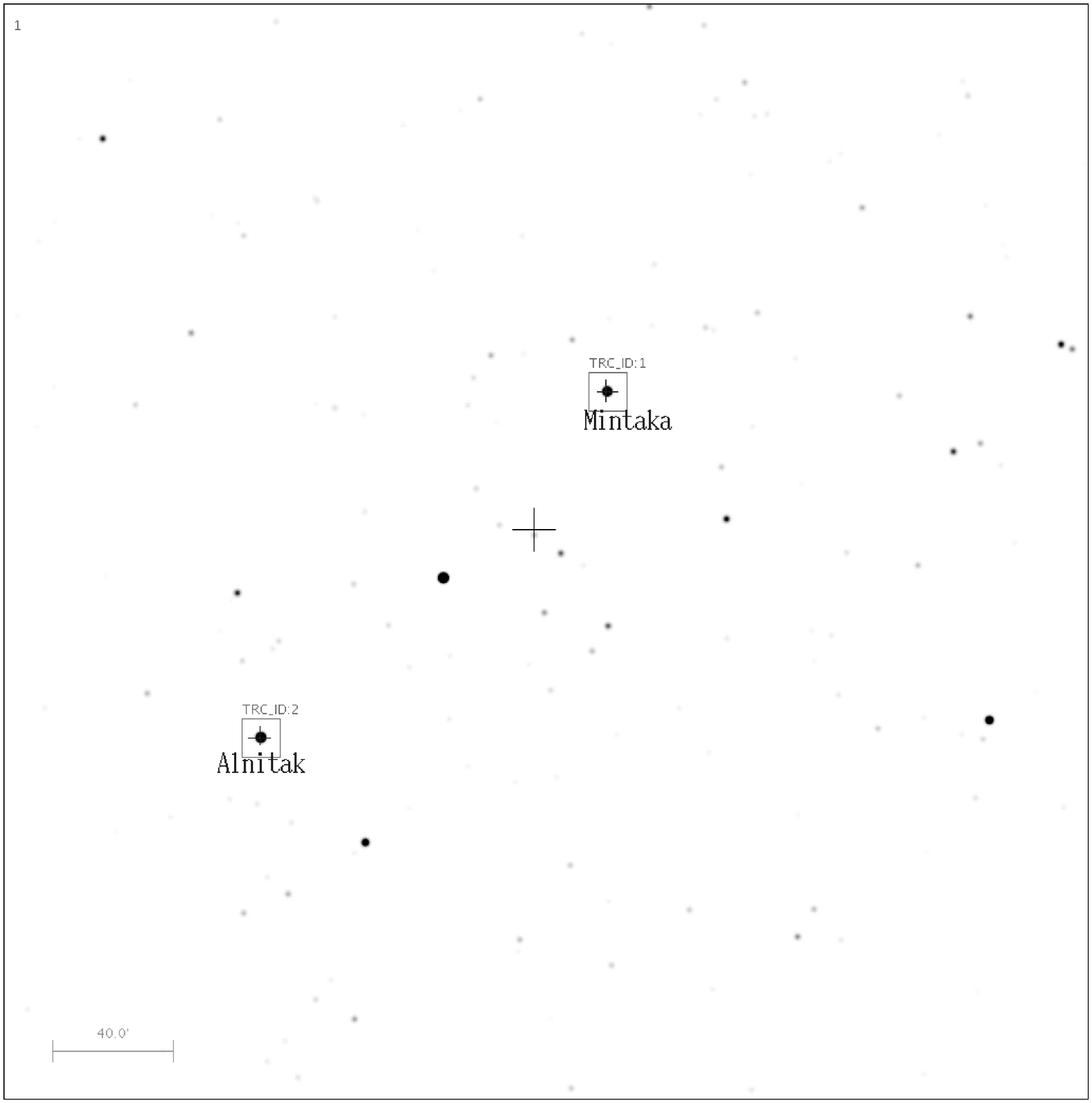}{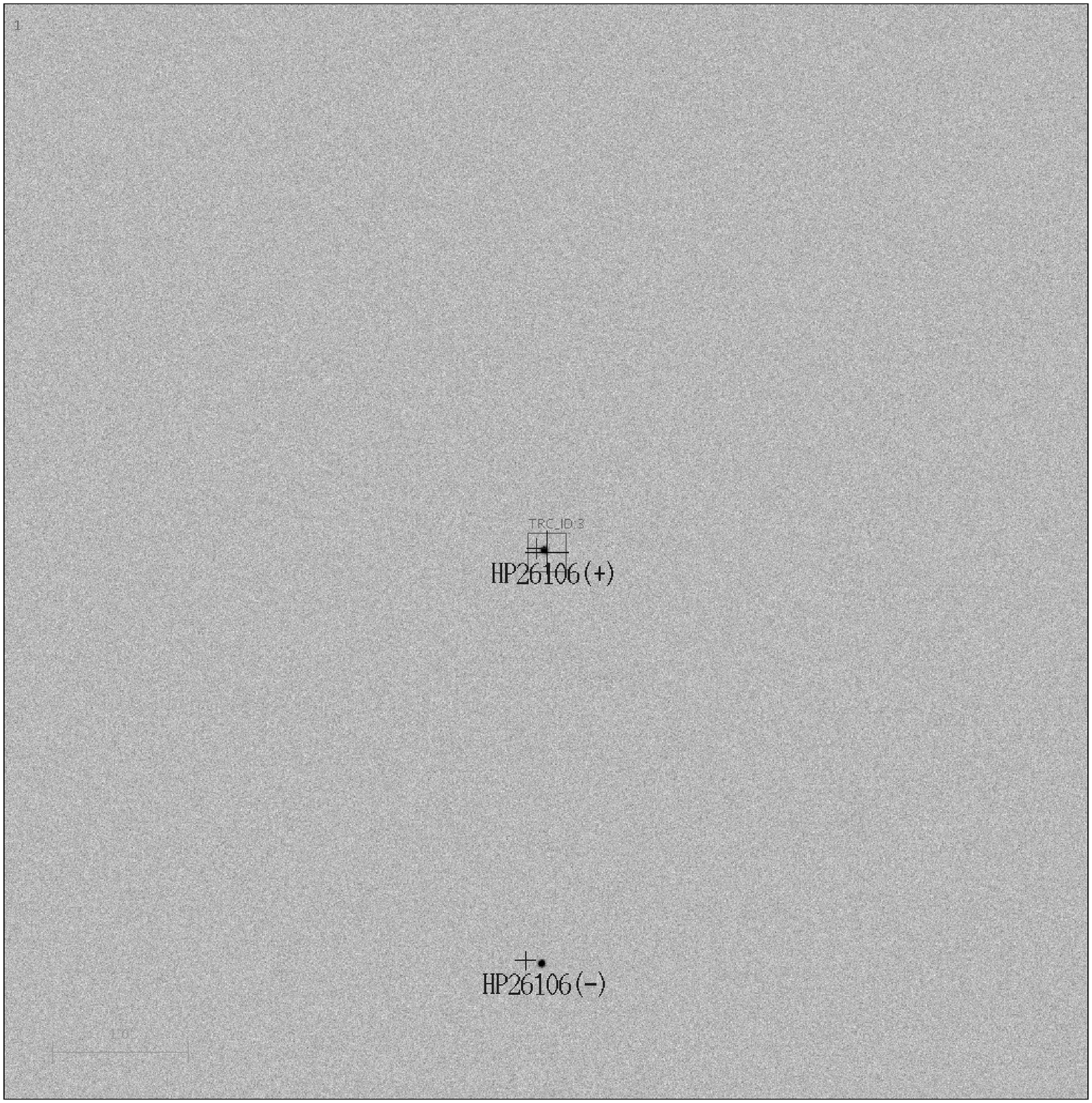}
\caption{Typical telescope setup for a single data acquisition.  Two
AOIs are defined on the WFI (left) to monitor
rotation.  A third is defined on the FPI (right) to monitor
transverse drift.  Computed positions
are superposed (for both chopped beams on the FPI), along with their names.
Mismatch represents several arcseconds of drift in absolute
pointing; depending on the precision required, a realignment may be
necessary.  Data shown is sky simulation
(Br\"uggenwirth, Gross, Nelbach, \& Shuping 2009)
with a thermal
dark current model.  The FPI is at $20^\circ$ C and the WFI is at $-40^\circ$
C.}
\label{fig:imagers}
\end{figure}

The MCCS uses a 2-3 stage approach to relate ERF to IRF, to $\la 1\arcsec$:
\begin{enumerate}
\item Rough-estimate ($\sim1^\circ$) ERF coordinates from avionics-reported
aircraft position and attitude and the TA's IRF and airframe attitudes.
\item Fine-tune using field recognition (PIXY; Yoshida 2000) on the Wide
Field Imager (WFI; 6$^\circ$ FOV).
\item Optionally, refine further using a 2DOF correction, based upon
two Focal Plane Imager (FPI; 0.57\arcsec pixels) acquisitions several degrees
apart.
\end{enumerate}
The operator monitors IRF$\mapsto$ERF with an overlay of calculated positions
of stars on imagers against sky images (see Fig.~\ref{fig:imagers}), and
executes the 2DOF refinement when mismatch drifts too large.

\section{Simple Pointing and Tracking}
The TA can be pointed in any directly observable coordinate system.
Areas of Interest (AOIs) can be defined on imagers; the TA
continually calculates AOI centroids.
This provides stability feedback to the TA, necessary because gyroscopes are not
sufficient to maintain pointing over hours.
A centroiding AOI can be attached to a position of interest; when activated,
the TA keeps the image fixed at the desired position, by modifying IRF to
cancel its motions inferred from the AOI.
The AOI need not coincide with the point of interest.

\section{Chopping}
A chop is a high-frequency square-wave oscillation of the secondary
mirror (SMA), synchronized electronically with imager acquisition to
subtract sky background.
Only the FPI and SI observe chopped images; other imagers do not 
share the optical path with the telescope.
A given FPI or SI pixel has two ERF coordinates since any images
taken are a superposition of two images with different centers;
the MCCS publishes both for key positions.
For pointing commands, positions can be modified by chop beam designation.
The SMA is commanded by a 2D center offset,
amplitude, and position angle, and a coordinate system.
SMA state is propagated through telescope motions by keeping its specification
constant in the coordinate system specified.
A ``3-point'' chop is also supported; this includes a pause at an intermediate
point (labelled ``0'').

\section{Nodding and Dithering}
\begin{figure}[!ht]
\epsscale{0.30}
\plotone{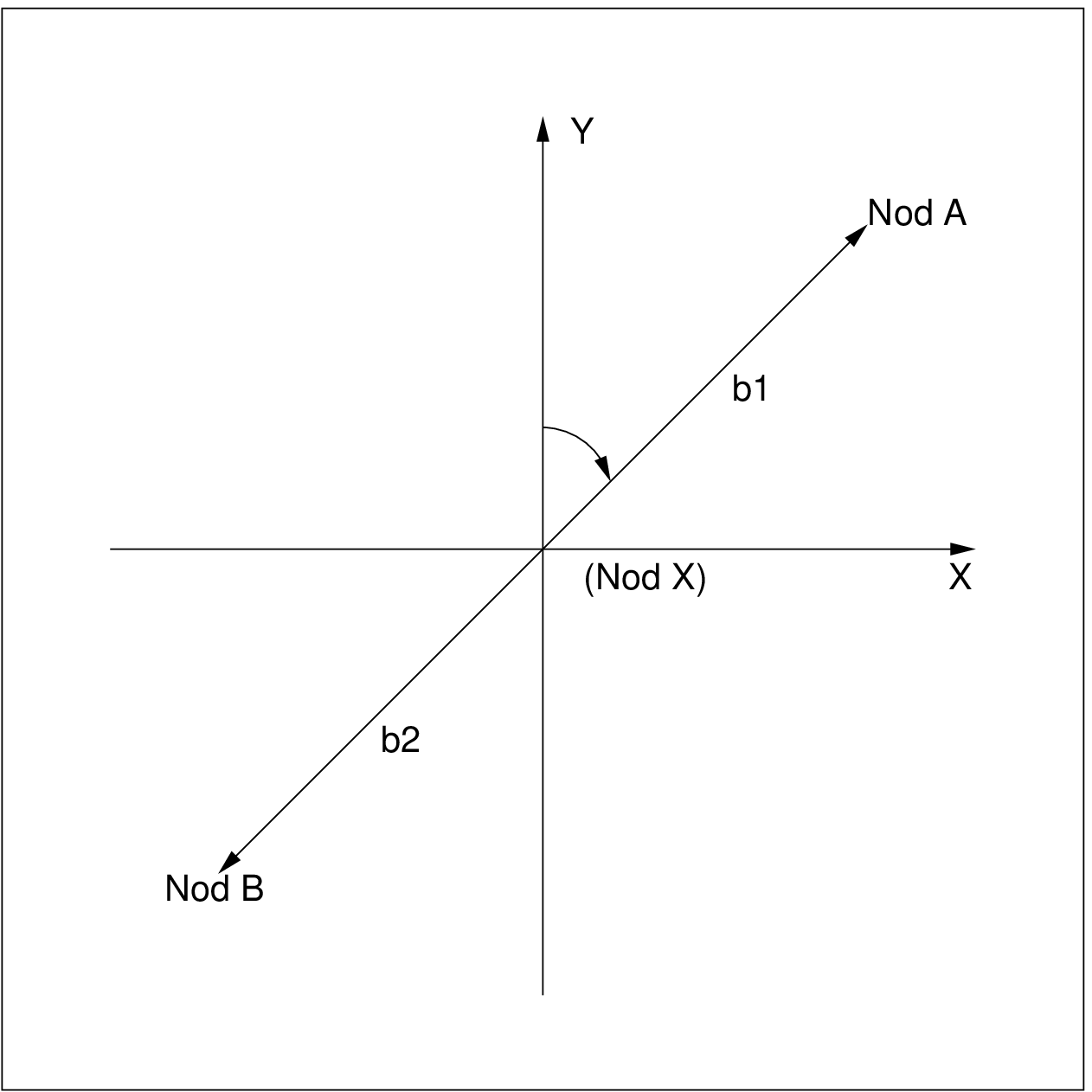}\hfil\plotone{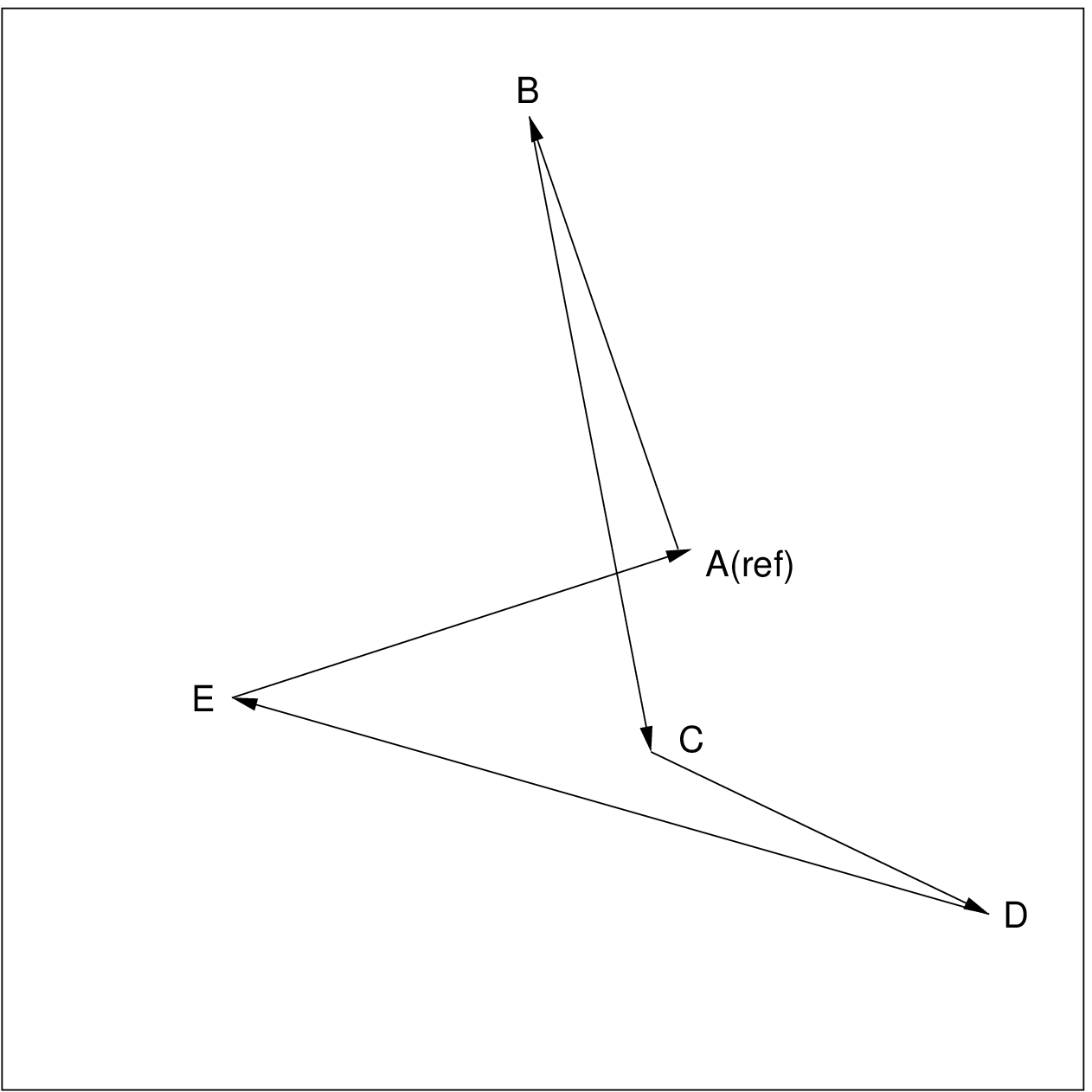}
\caption{Graphical description of nodding (left) and dithering (right).
Nods are specified in the same manner as chops, so they can be
matched unconditionally, without direct connections.  The center position
(``Nod X'')
is optional, and can be used to match a 3-point chop, which includes the
center position as well.  A formulation based
upon circular lists of relative pointings is used for dithers --
all positions are understood to be relative to the first.}
\end{figure}
A nod is a low-frequency motion of the whole telescope, usually relating
to an accompanying chop, to remove instrument backgrounds.
Nods are defined in the same manner as chops, without direct coupling,
and are propagated through telescope motions identically. 
Dithers are a generalization of nods to arbitrary-length
sequences, intended to image the same star on several different imager
pixels, in order to remove outliers such as cosmic ray hits or hot pixels.
The first position in a dither is
considered to be a ``reference,'' with all other positions defined relative
to it, for relocatability on the sky.
Chops, nods, and dithers have no direct coupling and can therefore
be mixed and matched in any combination.

\section{Managing Field Rotation}
TA motion is significantly constrained by clearance to the
airframe.  Thus, azimuth is mostly controlled by turning the
aircraft (Gross \& Shuping 2009).
Rotation against the sky can be held constant as the aircraft rotates
up to $5.6^\circ$.
As this rotation nears its limit, the telescope is periodically ``rewound''
manually to a new rotation angle, between data acquisitions.
During rewind, all building blocks are held constant in the
coordinate systems in which they were defined.

\section{Key Design Points and Tests}
In order to minimize development risk for the MCCS, several critical design
features have been included.  Pointing building blocks do not depend directly
on each other and state derives strictly from the TA,
allowing for independent development, testing, and operational use.
Much attention has
been given to abstraction and simplification, to aid in development of
testing procedures.
Each building block has a realistic operational ``scenario,''
defining detailed inputs, operator commands,
user-visible outputs, and intermediate calculations.

Coordinate systems present special problems, and are notoriously difficult
to orient correctly and keep that way; to that end, regression tests include:
\begin{itemize}
\item Diurnal rotation rate measurements with zero aircraft speed.
\item A ``home position'' at the equator on the vernal equinox,
with northbound heading and zero speed (enables manual calculations).
\item Modifying each coordinate parameter individually.
\item Composite tests (verifies calculations are in the correct order).
\item Corner cases (identifies quadrant problems).
\item Complete-system tests pointing to specific stars, using synthetic
images.
\end{itemize}
Some tests violate TA physical limits; these are evaluated against a simulator
(Br\"uggenwirth, Gross, Nelbach, \& Shuping 2009)

\section{Conclusion}
Placing a telescope on a moving platform, while still desiring to point
in absolute coordinates, presents a difficult challenge for software
development.  Eventual success of the SOFIA platform in this regard
depends upon careful abstraction and testing strategies for pointing
the telescope.

\end{document}